\documentclass[tightenlines,aps,
twocolumn,
showpacs,nofootinbib]{revtex4}

\usepackage{psfig,float}
\usepackage{amsmath,amsfonts,graphicx,bm}

\newcommand{\ep}{\epsilon}
\newcommand{\be}{\begin{equation}}
\newcommand{\ee}{\end{equation}}
\newcommand{\ba}{\begin{eqnarray}}
\newcommand{\ea}{\end{eqnarray}}

\begin{document}

\begin{titlepage}

\begin{flushright}
\vbox{
\begin{tabular}{l}
 SLAC-PUB-10360\\
 UH-511-1045-04\\
 hep-ph/0402280
\end{tabular}
}
\end{flushright}

\title{
Real radiation at NNLO: $e^+e^- \rightarrow 2$ jets through ${\cal O}(\alpha_{s}^{2})$
}

\author{
Charalampos Anastasiou\thanks{e-mail:babis@slac.stanford.edu}\vspace{-0.2cm}} 
\affiliation{
          Stanford Linear Accelerator Center,\\ 
          Stanford University, Stanford, CA 94309}
\author{Kirill Melnikov
        \thanks{e-mail: kirill@phys.hawaii.edu}\vspace{-0.2cm}}
\affiliation{Department of Physics and Astronomy,
          University of Hawaii,\\ 2505 Correa Rd., Honolulu, HI 96822}  
\author{Frank Petriello\thanks{frankjp@pha.jhu.edu}\vspace{-0.2cm}}
\affiliation{
Department of Physics, Johns Hopkins University, \\
3400 North Charles St., Baltimore, MD 21218
} 

\begin{abstract}
We present a  calculation of the
differential two jet cross section in $e^+e^-$ annihilation
through next-to-next-to-leading order in the strong coupling constant $\alpha_s$. 
The calculation is performed using a new method  for dealing with real radiation suggested 
recently by us in \cite{sector}. For the first time, 
the two jet event rate is computed directly, without 
any reference to the inclusive cross-section $e^+e^- \to {\rm hadrons}$.
We also calculate the  
energy distribution of the leading jet in $e^+e^- \to 2~{\rm jets}$ 
and find significant modifications of the shape of this 
distribution at NNLO.
\end{abstract}

\maketitle

\thispagestyle{empty}
\end{titlepage}

High-energy physics will begin to explore a new energy frontier 
when the Large Hadron Collider at CERN turns on in 2007.  
Our understanding of physics at very small distances will dramatically improve. 
However, a detailed investigation of the new physics we discover will require a 
careful study of Standard Model backgrounds, detector responses, and other 
similar issues.  Since many layers separate interesting  
physics from raw experimental data, a dedicated effort is required to fully 
utilize LHC results.  There have been significant advances towards this goal 
in the past few years; we now have an increased understanding of parton distribution 
functions and jet algorithms, improved Monte Carlo event generators, methods 
for automating next-to-leading order calculations with large 
number of external legs and, finally, new technology for next-to-next-to-leading order (NNLO)
computations.  

NNLO calculations are certainly not required for all processes at the LHC or existing colliders; however, 
there are a few situations in which NNLO calculations are highly desirable.
These include processes for which the one-loop corrections are abnormally large (e.g. the 
production of the SM Higgs boson at hadron colliders
\cite{Harlander:2002wh}) 
or for measurements in which high experimental precision is either achieved 
(e.g. the $\alpha_s$ determination 
from the three-jet event rate in $e^+e^-$ annihilation \cite{Bethke:2000ai} or 
the $W$ mass measurement at the Tevatron)
or expected  (e.g. $W$ and $Z$ boson production at the LHC
\cite{lum}).  These calculations should 
also inform us how accurate NLO calculations really are, beyond the standard checks 
of stability with respect to renormalization and factorization 
scales variations.  We should learn to estimate the significance 
of NNLO corrections without performing the calculations, given the required 
precision for an observable and the kinematic regions in which 
it is measured.  For this purpose, exclusive 
NNLO calculations are needed, since 
experimental cuts on the final state can have a strong 
impact on the convergence of perturbative expansion. 
Unfortunately, not a single calculation of a fully 
differential QCD observable at NNLO has been performed, either for 
lepton or hadron colliders.

In this Letter we remedy this situation and
present the calculation of the two jet 
cross-section in $e^+e^-$ annihilation
at NNLO in perturbative QCD.  Although jets and their properties
have been studied very extensively at lepton colliders
\cite{catani}, we believe that such calculation is important for the
following reasons: 1) it is the first-ever calculation of a
{\it fully differential} observable at NNLO; 2) although 
the total rate for two jet events in $e^+e^-$ 
annihilation is known through NNLO from indirect calculations, 
our results for distributions in two jet events are new; 
3) this calculation is possible because of a new 
method we recently suggested for handling real radiation in 
hard processes \cite{sector}; it is important to demonstrate 
its efficiency by applying it to a non-trivial example.

There is a strong correlation between 
the complexity of higher order calculations and the level of 
exclusiveness desired.  Traditionally, it was thought that calculations at higher 
orders are difficult because of multi-loop integrals.  A significant effort therefore 
went into developing flexible, easily automated methods for performing higher 
loop computations \cite{tkachov}.  As a result, calculations up 
to two loops are no longer prohibitively difficult; for example, the two-loop virtual corrections
for $1 \to 3$ and all partonic $2 \to 2$  processes at hadron colliders 
have been computed \cite{2loop}.  
Surprisingly, the major obstacle in obtaining differential results at higher orders 
are tree-level processes with additional final-state partons.  While it is very easy to write down the 
corresponding matrix element, it is not possible to integrate it numerically over 
the restricted (exclusive) phase-space  without first extracting 
the singular structure of the integrand in the soft and collinear limits.  
Analytic integrations also become extremely difficult because of the 
arbitrariness of final-state cuts.  This problem has been successfully solved at NLO using both the slicing and dipole 
subtraction methods \cite{slice,dipole}.  Attempts have been made recently 
to generalize the dipole formalism to NNLO \cite{duck};  
so far, they have not completely succeeded.  It is therefore productive 
to look for an alternative method of  dealing with multi-particle final 
states in the presence of arbitrary constraints on their phase space.

What are the ideal features of such a method?  Given the complexity of 
higher order calculations, it should satisfy the following requirements:
1) the  singularities should be extracted 
in an algorithmic fashion;
2) the method  should be easy to automate; 
3) it should be easily generalizable, at least in principle, 
to arbitrary numbers of partons in the final state;
4) the method should work efficiently in the presence of arbitrary 
constraints on the final state;
5) it should lead to a fast and accurate numerical evaluation of physical quantities.

We have proposed such a method recently in \cite{sector}.  
We now briefly describe its salient features.  Consider a perturbative
tree-amplitude ${\cal M}$ with $n$ particles in the final state. 
Its contribution to the differential cross-section can be written as
\vspace{-0.2cm}
\be
\vspace{-0.2cm}
{\rm d}\sigma^{(n)} = \int {\rm d} \Gamma_n |{\cal M}|^2 J(\{p_i\}), 
\label{eq1}
\ee
where ${\rm d}\Gamma_n$ denotes the $n$-particle phase-space and 
$J(\{p_i\})$ imposes restrictions on the final state (e.g. the jet algorithm) 
which define the experimentally observed process. Throughout this 
Letter we use dimensional regularization (with $d=4-2\epsilon$ dimensions) 
for both infrared and collinear divergences.
If all particles are well separated (resolved), $|{\cal M}|^2$ is 
finite; however, when the integration in Eq.(\ref{eq1}) 
is attempted, there are divergences associated with soft and 
collinear kinematic configurations.  A direct numerical integration 
of Eq.(\ref{eq1}) is therefore not possible.  

In Ref.~\cite{sector} we demonstrated that by mapping the invariant masses 
onto algorithmically chosen new 
integration variables, it is possible to extract the $\epsilon$ poles 
explicitly. We then obtain an expansion in $\epsilon$,
\vspace{-0.15cm}
\be
\vspace{-0.15cm}
{\rm d}\sigma^{(n)} = \sum \limits_{k=0}^{2(n-2)}
\frac{{\rm d}F_k}{\epsilon^{2(n-2)-k}}+ {\cal O}(\ep),
\label{eq2}
\ee
where the coefficients ${\rm d}F_k$ are well-defined, $\ep$-independent multi-dimensional 
integrals for a generic function ${\cal J}$.  It needs to 
be specified only at the stage of numerical evaluation.  This allows us to derive 
results for arbitrary jet algorithms and experimental observables. 

It is relatively easy to derive such an expansion at NLO, where at most one 
parton can become unresolved. In this case, trivial 
mappings~\cite{sector} of the phase-space variables onto variables with range from 0 to 1 
produce integrals with the following singular structure: 
\vspace{-0.15cm}
\be
\vspace{-0.15cm}
 I_1 =  \int \limits_{0}^{1} {\rm d}x {\rm d}y~x^{\ep-1} y^{\ep-1} 
~J(x,y).
\ee
All singularities in the above integral can be extracted by writing 
$x^{\ep-1} = \delta(x)/\ep + [1/x]_+ +\ep [\ln(x)/x]_+ +...$, for 
both $x$ and $y$ in the integrand. 

Beyond NLO, two or more partons may become unresolved, which gives rise to 
a more complicated structure of overlapping singularities. Typically, we find integrals similar to 
\vspace{-0.15cm}
\be
\vspace{-0.15cm}
I_2 = \int \limits_{0}^{1} {\rm d}x {\rm d}y~ 
~\frac{x^{\ep} y^{\ep}}{(x+y)^2}~J(x,y).
\label{eq3}
\ee
The procedure described above does not work because the singularities 
are not factorized.  To solve this problem, we apply the technique of sector 
decomposition~\cite{decomp}; we divide the integration region in 
Eq.(\ref{eq3}) into patches with a definite ordering of the integration 
variables ($x <y$ and $y < x$) and reweight all variables in each patch 
so that the integrations again range from 0 to 1.  This leads to factorization of the 
singular limits.  This procedure can be 
completely automated.  The same method should, in principle, work for any number 
of particles in the final state, both massless and massive, and for any restrictions 
on the final-state phase space.

For $e^+e^- \to 2~{\rm jets}$ through NNLO, the largest multiplicity 
of particles in the final state is four (e.g. $e^+e^- \to q\bar q gg$).
A parameterization of the $1 \to 4$ particle phase-space in terms of five 
independent variables which is suitable for sector decomposition and 
extracting infrared divergences was given in \cite{sector}.  
In the same reference, we gave a more detailed description of 
the method, and considered a number of relatively simple examples.  
In this Letter we apply the method to a fully realistic and non-trivial 
problem --  the calculation of the $e^+e^- \to 2~{\rm jets}$ cross section
at NNLO.  Traditionally, the inclusive 2-jet rate is calculated at NNLO 
indirectly, by first computing 
the total inclusive cross-section for $e^+e^- \to {\rm hadrons}$ 
and then subtracting from it the $e^+e^- \to 4~{\rm jets}$ 
and $e^+e^- \to 3~{\rm jets}$ cross-sections at LO and NLO, respectively.  
This paper presents the first direct calculation of the 2-jet rate at NNLO.  
Using our method, we can also obtain differential results at NNLO, 
which can not be derived indirectly.  We illustrate this by computing 
the energy distribution of the leading jet in $e^+e^- \to 2~{\rm jets}$ 
through NNLO.

The cross-section for $e^+e^-$ annihilation into hadrons 
through order ${\cal O}(\alpha_s^2)$ can be written as
\ba
&& \sigma = \sigma_0 
\left ( \delta_{j,2} + \left( \frac{\alpha_s}{\pi} \right ) 
\left ( C_1^{(2)}\delta_{j,2} + C_1^{(3)} \delta_{j,3} \right )
\right .\nonumber \\
&& \left. + \left ( \frac{\alpha_s}{\pi} \right )^2 
\left ( C_2^{(2)}\delta_{j,2} + C_2^{(3)} \delta_{j,3}
 + C_2^{(4)} \delta_{j,4} \right ) \right ),
\ea
where $\sigma_0 = 4\pi \alpha_{\rm QED} \sum_q Q_q^2/s$ is the 
tree level cross-section for $e^+e^- \to q \bar q$,
$\sqrt{s}$ is the center of mass energy,
$\alpha_s  = \alpha_s(s)$ is the ${\overline  {\rm MS}}$ 
QCD coupling constant and the coefficients $C_{i}^{(j)}$
describe  jet cross-sections in various 
orders in perturbation theory, as indicated by the Kronecker
symbols. From inclusive  calculations of the cross-section \cite{Dine:1979qh}, we find
\ba
&& C_1^{(2)} + C_1^{(3)}= 1, ~~~~~C_2^{(2)}+ C_2^{(3)} + C_2^{(4)} =
\label{eq6}
\\
&& \frac{365}{24}-11\zeta(3)-\left (
\frac{11}{12}-\frac{2\zeta(3)}{3} \right )N_f
\approx 1.99 - 0.115 N_f, \nonumber
\ea
where $N_f$ is the number of massless fermion flavors.

An important goal of this Letter is the calculation of the 
coefficient $C_2^{(2)}$, the NNLO correction to the 
two-jet production rate.  For this, we need 
the two-loop virtual correction to $e^+e^- \to q \bar q$, the one-loop 
correction to the $e^+e^- \to q \bar q g$ process, and 
the tree level processes $e^+e^- \to q\bar q gg $ 
and $e^+e^- \to q \bar q q_1 \bar q_1$.  We also require the coupling 
constant renormalization of the NLO result.  At order ${\cal O}(\alpha_s^2)$, 
all of these processes contain divergent contributions to the two 
jet cross-sections; the highest singularity is 
$1/\ep^4$. The singularities cancel when individual contributions 
are combined to form a physical observable.

The two-loop virtual corrections to $e^+e^- \to q \bar q$ are 
well-known~\cite{vanNeerven:1985xr}. 
We have outlined  above how the tree-level four parton final state 
is handled in our approach.  We note that a global parameterization 
of the four particle phase space, which we used in \cite{sector} for the $N_f$ terms, 
leads to large analytic expressions which are difficult to evaluate 
numerically.  We found it much more convenient to select a different parameterization 
for the invariant masses in each individual term, thereby reducing 
the number of sector decompositions required to extract the singularities.
This choice of parameterization can be done automatically 
once the basis topologies appearing in the matrix element are identified.  
With this clever choice of parameterization, the size of the computer 
code for the fully differential NNLO $e^+e^- \to 2~{\rm jets}$ process
is not much larger than what we have found in simpler
examples in \cite{sector}.  The required CPU time is also 
not very large; to achieve the precision on the jet rates presented in this paper, 
about four hours are needed on a PC with a $3~{\rm GHz}$ Pentium 4 processor.

We now briefly comment on the calculation of the 
one-loop corrections to the $q \bar q g$ final state.  
It might seem that a different technique is needed 
to handle this contribution, since a virtual loop integration is involved.
However, this is not the case \cite{sector}.  Once the virtual 
loop integrals are expressed through Feynman parameters, 
they can be treated identically to phase-space integrals.  
We found it convenient to express them through standard hypergeometric 
functions, and use the one-dimensional integral representation for the 
hypergeometric functions, together with the three-parton phase space 
parameterization.  Although this procedure is not necessary, it is useful because 
it provides an economical input for sector decomposition.

Since our approach to the problem is numerical, 
including the cancellation of $1/\ep$ poles, we must consider 
issues of numerical accuracy. The simplest check is the comparison 
of the direct and indirect evaluations of the total two-jet event 
rate.  The indirect result is obtained by taking the difference
between the ${\cal O}(\alpha_{s}^{s})$ contribution to the inclusive cross-section,
given in Eq. (\ref{eq6}), and subtracting from it the four-jet 
cross-section at LO and the three-jet cross-section at NLO. 
Both of these quantities are computed in our code.  
We use the JADE algorithm \cite{jade} to identify jets in the final 
state. However, the jet definition is an independent subroutine 
in our code that can be trivially changed if desired.  
Choosing the jet separation parameter for the JADE algorithm 
$y_{\rm cut} = 0.1$, we obtain
\be
C_2^{(2),{\rm indirect}} = \left(-49.2\pm0.4\right)+\left(1.7974\pm0.0011\right)N_f,
\label{eq7}
\ee
where the errors denote our integration uncertainties for the 3 and 4 jet 
cross sections.
A direct computation of the same quantity yields
\begin{eqnarray}
C_2^{(2)} &=& \frac{10^{-6}}{\ep^4}
+\frac{10^{-4}}{\ep^3}
+\frac{10^{-3}}{\ep^2} + \frac{\left(-4\pm4\right) \times 10^{-2}}{\ep} 
\nonumber \\ & & + \frac{\left(-0.3\pm4\right)\times 10^{-4}}{\ep}N_f
+ \left(-49.8\pm0.4\right)\nonumber \\ & & +\left(1.798\pm0.002\right)N_f \,\, .
\label{eq8}
\end{eqnarray}
We have included the integration errors found during an actual run for 
the $1/\ep$ poles to demonstrate the level of cancellation; the 
magnitudes indicated for the higher poles are typical of results 
found using our code.  Comparing Eq.(\ref{eq7}) and Eq.(\ref{eq8}), 
we conclude that our program provides a precision on the finite part of 
the NNLO correction to the two-jet rate better than 1\%.  We also conclude that 
our numerical cancellation of $1/\ep$ poles works very efficiently.  
These features do not change significantly when the 
jet separation parameter $y_{\rm cut}$ is varied.

\noindent
\begin{figure}[t]
\centerline{
\psfig{figure=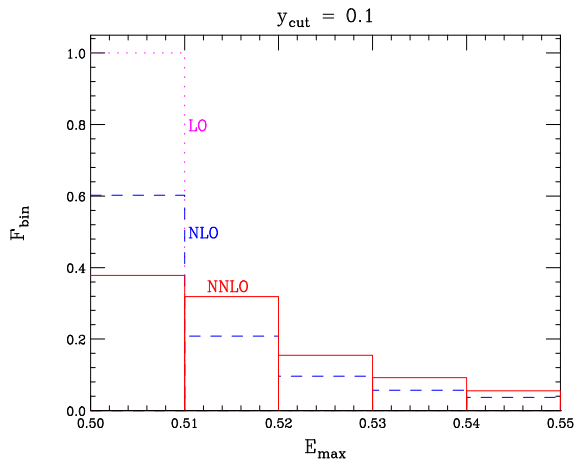,height=6.0cm,width=7.5cm,angle=0}}
\vspace*{0.5cm}
\caption{Bin-integrated energy distribution for $y_{\rm cut} = 0.1$.  
The fractions of events in each energy bin are shown.  The 
dotted histogram denotes the LO result, the dashed histogram the NLO result, and the solid histogram 
the NNLO result.}
\label{enplots}
\end{figure}

Our approach permits us to also compute 
differential distributions in addition to the total rate.  
As an example, we present below the energy distribution of the leading jet
in two jet events at NNLO. At leading order, this 
distribution is simple; since two massless quarks are produced, each jet 
contains half of the total energy.  
The distribution becomes more interesting at NLO, when it becomes 
possible for one of the jets to have an invariant mass different 
from zero.  At NNLO, configurations when the invariant masses 
of both jets are different from zero appear for the first time.  
We compute this distribution by a simple modification of the jet function;
after an event is identified as a two jet event, 
the energies of the two jets are computed and 
the jet with the largest energy is identified. This number is 
then stored in the appropriate bin of a histogram. 
The corresponding bin-integrated energy distribution 
is shown in  Fig. 1 for $y_{\rm cut} = 0.1$, $N_f=5$, and $\alpha_s = 0.121$.  
The distribution is significantly distorted by NNLO QCD corrections; 
the corrections are large for this $y_{\rm cut}$, and many 
situations that look like three and four jet events are 
identified as two jet events.  Smaller $y_{\rm cut}$ choices lead to large 
logarithms in the perturbative expansion that invalidate the fixed order 
result.

In conclusion, we have presented the first calculation 
of the NNLO corrections to a fully differential observable in QCD.  We have 
demonstrated our approach using the non-trivial example of 
$e^+e^- \rightarrow$ 2 jets.  We have computed the NNLO corrections to the 
energy distribution of the two jets in $e^+e^-$ annihilation, and 
have shown that the shape of the distribution changes when the NNLO 
corrections are included.  Our method allows other phenomenologically interesting distributions 
in 2-jet events to be easily computed; these will be discussed elsewhere.  Since our approach is 
fully numerical, we have presented convincing evidence that 
reasonable precision and control of numerical stability can be achieved.
The method we developed for this calculation is quite flexible; it generalizes 
straightforwardly to an arbitrary number of partons in the final 
state, both massive and massless.  Given unlimited computing resources, 
it provides a complete solution to the problem of real radiation at higher 
orders in perturbative QCD.  In practice, significant effort 
and some ingenuity will be required to apply it to more complicated processes 
of direct phenomenological relevance.  We look forward to this challenge.

K. M. thanks  the KITP, UC Santa Barbara for its hospitality.  
C. A. thanks the Johns Hopkins University for its hospitality.  
This research was supported by the US Department of Energy  under contracts
DE-AC03-76SF0515, DE-FG03-94ER-40833 and 
the Outstanding Junior Investigator Award DE-FG03-94ER-40833, and`
by the National Science 
Foundation  under contracts P420D3620414350, P420D3620434350 
and partially under Grant No. PHY99-07949.


\end{document}